\documentstyle[12pt]{article}
\date{}
\newcommand{\half}{\mbox{\small$\frac{1}{2}$}}

\newcommand{\calH}{{\cal H}}
\newcommand{\be}{\begin{equation}}
\newcommand{\ee}{\end{equation}}
\newcommand{\bea}{\begin{eqnarray}}
\newcommand{\eea}{\end{eqnarray}}
\newcommand{\mye}{\mbox{e}}
\newcommand{\ph}{\phi}
\def\rellow#1#2{\mathrel{\mathop{\kern 0pt #1}\limits_{#2}}}

\newcommand{\str}{\rule{0ex}{2.7ex}}  
\newcommand{\strr}{\rule{0ex}{4.0ex}}  
\newcommand{\tabhline}{\\[0.3ex] \hline \str}

\newcommand{\Um}[1]{U_{#1,\mu}}
\newcommand{\Un}[1]{U_{#1,\nu}}
\newcommand{\Ut}[1]{U_{#1,\tau}}
\newcommand{\Tr}{\mbox{Tr}}
\newcommand{\foN}{ \mbox{ \small$\frac{1}{N}$ } }
\newcommand{\EW}[1]{ \langle #1 \rangle }
\newcommand{\vn}{\vec n}
\newcommand{\vs}{\vec \sigma}
\newcommand{\calP}{{\cal P}}
\newcommand{\Lamt}{ \Lambda_t^{\tau} }
\newcommand{\erfc}{\mbox{erfc}}
\newcommand{\delH}{\Delta\calH}

\title{\vbox{\vspace{.1mm}}
Kinematics of Multigrid Monte Carlo}
\author{ \vbox{\vspace{7mm}}
   {\bf Martin Grabenstein$^{1}$ and Klaus Pinn$^{2}$} \\[6mm]
$^1\,$II. Institut f\"ur Theoretische Physik, Universit\"at Hamburg,
      \\[-1mm]
      Luruper Chaussee 149, D-2000 Hamburg 50, Germany,
      \\[-1mm]
      {\tt \small  bitnet I$\!\, \not \! 0$2GRA@DHHDESY3 }
      \\[4mm]
$^2\,$Institut f\"ur Theoretische Physik I, Universit\"at M\"unster,\\
      [-1mm]
      Wilhelm-Klemm-Str.\ 9, D-4400 M\"unster, Germany,
      \\[-1mm]
      {\tt \small bitnet PINN@DMSWWU1A    }           \\[18mm]
      { preprint DESY 92-094, July 1992}           \\[4mm] }
\begin{document}
\maketitle \vfill
\begin{abstract} \normalsize
We study the kinematics of multigrid Monte Carlo algorithms by means of
acceptance rates for nonlocal Metropolis update proposals.  An
approximation formula for acceptance rates is derived.  We present a
comparison of different coarse-to-fine interpolation schemes in free
field theory, where the formula is exact.
The predictions of the approximation formula for several interacting
models are well confirmed by Monte Carlo simulations.
The following rule is found: For a critical
model with fundamental Hamiltonian $\calH(\ph)$,
absence of critical slowing down can only be expected if the
expansion of $\langle \calH (\ph + \psi) \rangle$ in terms of the shift
$\psi$ contains no relevant (mass) term.
We also introduce a multigrid update procedure for
nonabelian lattice gauge theory and study the acceptance rates for
gauge group $SU(2)$ in four dimensions.
\end{abstract} \vfill
\thispagestyle{empty}

\newpage
 \pagenumbering{roman}
 \tableofcontents

\newpage
\pagenumbering{arabic}


\section{Introduction}
\label{SECintro}

Monte Carlo simulations of critical or nearly critical statistical
mechanical systems with local algorithms suffer from critical slowing
down (CSD).  Roughly speaking, the autocorrelation time in the Markov
chain behaves like $\tau \sim \xi^z$ in the vicinity of a
critical point, where $\xi$ denotes the correlation
length, and $z$ is the dynamical critical exponent.  For conventional
local algorithms, $z \approx 2$.  For accelerated local algorithms such
as overrelaxation or the optimized hybrid Monte Carlo algorithm, one can
sometimes achieve $z \approx 1$ \cite{over,hybrid}.  To overcome the
problem of CSD, various nonlocal Monte Carlo algorithms have been
developed.

Cluster algorithms \cite{cluster} are successful in overcoming CSD for a
large class of models.  The alternative is multigrid Monte Carlo
\cite{multigrid,sokalprl,cargmack}.  In this paper, every algorithm that
updates stochastic variables on a hierarchy of length scales is called
multigrid Monte Carlo algorithm.  There are models where no successful
cluster algorithms have been found whereas multigrid Monte Carlo
algorithms work \cite{sunsun,hmsun}.

Presently, the only generally applicable method to study algorithms for
interacting models is numerical experiment.  It is however important to
have some theoretical understanding that helps to predict which
algorithms will have a chance to overcome CSD in simulations of a given
model.  As a contribution to the research in this direction we here
present a study of the kinematics of multigrid Monte Carlo algorithms
\footnote{Parts of this paper are published in short form in
\cite{schonda}}.
With kinematics we here mean the study of the scale (block size)
dependence of the Metropolis acceptance rates for nonlocal update
proposals.  We do not address the much more ambitious problem of
analytically investigating the full dynamical critical behavior of the
stochastic processes involved.  Our analysis is nonetheless of
relevance because sufficiently high acceptance rates
are necessary for multigrid Monte Carlo procedures to overcome CSD.

We derive an approximation formula for the block size dependence of
acceptance rates for nonlocal Metropolis updates.  The influence of the
coarse-to-fine interpolation kernel (shape function) on the kinematics
in free field theory, where the formula is exact, is investigated in
detail.

The formula is then applied on several interacting
models and turns out to be a very
good approximation.  We find necessary criteria for a given multigrid
algorithm to eliminate CSD: For a critical
model with a fundamental Hamiltonian $\calH(\ph)$
absence of CSD can only be expected if the expansion
of $\langle \calH (\ph + \psi) \rangle$ in terms of the shift $\psi$
contains no relevant term (mass term).

There is an urgent demand for accelerated Monte Carlo algorithms
in lattice gauge theory. The present state-of-the-art algorithm
is overrelaxation \cite{overgauge}.
However, effort was also spent in developing nonlocal algorithms
for gauge theories. An efficient cluster algorithm
was found for $SU(2)$ gauge theory at finite temperature,
however only in the special case $N_t=1$ \cite{su2fint}.
For a recent cluster algorithm approach to $U(1)$ gauge theory
see \cite{sinclair}.
Multigrid algorithms for $U(1)$ gauge models were
introduced and studied in two and four dimensions
\cite{laursen,newgauge}.
A different but related nonlocal updating scheme in the
abelian case is the multiscale method \cite{adler}.

In this paper, we propose a multigrid algorithm for nonabelian gauge
theory and analyze its kinematics.  Our approximation formula turns out
to be very reliable also in this case and allows for a prediction of
acceptance rates for a large class of nonlocal updates.

We believe that the proposed algorithm should be able to
accelerate the local Monte Carlo dynamics at least by a constant
factor.

This paper is organized as follows:  In section \ref{SECunigrid} we
introduce multigrid Monte Carlo algorithms.
Section \ref{SECapprox} contains the derivation of our
approximation formula for acceptance rates.
Several coarse-to-fine-interpolation
kernels are discussed in section \ref{SECkernels}. In
section \ref{SECfree} the
acceptance rates in free field theory are examined in detail.  The
kinematical analysis for the Sine Gordon, XY and $\phi^4$ models is
presented in section \ref{SECappl}.  In section \ref{SECgauge} we
propose a multigrid procedure for nonabelian gauge theories and analyze
its kinematics. A summary is given in section \ref{SECsummary}.

\section{Multigrid Monte Carlo algorithms}

\label{SECunigrid}

We consider lattice models with partition functions
\be\label{PARTI}
Z= \int \prod_{x \in \Lambda_0} d\ph_x \, \exp(-\calH(\ph)) \,
\ee
on cubic $d$-dimensional lattices $\Lambda_0$ with periodic
boundary conditions. The lattice spacing is set to one. We
use dimensionless spin variables $\ph_x$.
An example is single-component $\phi^4$-theory, defined by the
Hamiltonian
\be
\calH(\ph) = \half(\ph, -\Delta \ph)
           + \frac{m_o^2}2 \sum_x \ph_x^2
           + \frac{\lambda_o}{4!} \sum_x \ph_x^4 \, ,
\ee
where
\be \label{DELTA}
(\ph, -\Delta \ph) = \sum_{<x,y>} (\ph_x - \ph_y)^2 \, .
\ee
The sum in eq.\ (\ref{DELTA}) is over all nearest neighbor pairs
in the lattice
\footnote{The definitions for lattice gauge theory will be introduced
in  section \ref{SECgauge}}.

A standard algorithm to perform Monte Carlo simulations of a model
of the type defined above is the local Metropolis algorithm:
One visits in a regular or random order the sites of the lattice
and performs the following steps: At site $x_o$, one
proposes a shift
\be
\ph_{x_o} \rightarrow  \ph_{x_o}' = \ph_{x_o} + s \, .
\ee
The configuration $\{ \ph_x \}$ remains unchanged for $x \neq x_o$.
$s$ is a random number selected according to an a priori
distribution $\rho(s)$ which is symmetric with respect to
$s \rightarrow - s$. E.g., one selects $s$ with uniform
probability from an interval $[-\varepsilon,\varepsilon]$.
One then computes the change of the Hamiltonian
\be
\Delta \calH = \calH(\ph')-\calH(\ph) \, .
\ee
Finally the proposed shift is accepted with
probability $\min [1,\exp(-\Delta \calH)]$. Then one proceeds
to the next site.

The local Metropolis algorithm suffers from CSD when
the correlation length in the system becomes large: long
wavelength fluctuations cannot efficiently be generated
by a sequence of local operations.
It is therefore natural to study nonlocal generalizations
of the update procedure defined above.

Consider the fundamental lattice $\Lambda_0$ as divided in
cubic blocks of size $l^d$. This defines a block lattice
$\Lambda_1$. By iterating this procedure one obtains a whole
hierarchy of block lattices $\Lambda_0, \Lambda_1, \dots, \Lambda_K$
with increasing lattice spacing.
This hierarchy of lattices is called multigrid.

Let us denote block lattice points in $\Lambda_k$ by $x'$.
Block spins $\Phi_{x'}$ are defined on block lattices
$\Lambda_k$. They are averages of the fundamental field $\ph_x$
over blocks of side length $L_B=l^k$:
\be\label{average}
\Phi_{x'} = L^{(d-2)/2}_B \, L^{-d}_B  \sum_{x \in x'} \ph_x \, .
\ee
The sum is over all points $x$ in the block $x'$.
The $L_B$-dependent factor in front of the average
comes from the fact that the corresponding dimensionful
block spins are measured in units of the
block lattice spacing:
A scalar field $\ph(x)$ in $d$ dimensions has canonical
dimension $(2-d)/2$. Thus $\ph(x) = a^{(2-d)/2}\ph_x$, where
$a$ denotes the fundamental lattice spacing. Now measure the
dimensionful block spin $\Phi(x')$ in units of the
block lattice spacing $a'$:
$\Phi(x') = a'^{(2-d)/2}\Phi_{x'}$ , with $a' = a L_B$.
If we average in a natural way
$\Phi(x')= L_B^{-d}\sum_{x \in x'}\phi(x)$ and return to
dimensionless variables, we obtain eq.\ (\ref{average}).

A nonlocal change of the configuration $\ph$ consists
of a shift
\be \label{nonlocal}
\ph_x \rightarrow \ph_x + s \, \psi_x \, .
\ee
$s$ is a real parameter, and the
``coarse-to-fine interpolation kernel'' (or shape function)
$\psi_x$ determines the shape of the nonlocal change. $\psi$ is
normalized according to
\be\label{normpsi}
L^{-d}_B \sum_{x \in x'} \psi_x = L^{(2-d)/2}_B \delta_{x',x_o'} \, .
\ee
Note that by the nonlocal change (\ref{nonlocal}), the block spin
is moved as $\Phi_{x'} \rightarrow \Phi_{x'} + s $
for $x'=x_o'$, and remains unchanged on the other blocks.
The simplest choice of the kernel
$\psi$ that obeys the constraint
(\ref{normpsi}) is a piecewise constant
kernel: $\psi_x = L^{(2-d)/2}_B$, if $x \in x_o'$, and $0$ otherwise.
Other kernels are smooth and thus avoid
large energy costs from the block boundaries. A systematic
study of different kernels will be given in section
\ref{SECkernels} below.

The $s$-dependent Metropolis acceptance rate for such
proposals is given by
\be\label{omega}
\Omega(s) = \bigl<
\min \lbrack 1 , \exp( - \Delta \calH) \rbrack \bigr> \, .
\ee
Here, $\bigl< (.)\bigr>$ denotes the expectation value in the
system defined by eq.\ (\ref{PARTI}). Furthermore,
\be
\Delta \calH = \calH(\ph + s\psi)-\calH(\ph) \, .
\ee
$\Omega(s)$ can be interpreted as the acceptance rate for
shifting block spins by an amount of $s$, averaged over
a sequence of configurations generated by a Monte Carlo simulation.
Note that $\Omega(s)$ does not depend on the algorithm that we
use to compute it.
$\Omega(s)$ is a useful quantity when one wants to know how efficiently
updates with increasing nonlocality (i.e.\ increasing
block size $L_B$) can be performed. Of course,
different choices of the kernel $\psi$ result in different acceptance
rates.

In actual Monte Carlo simulations, $s$ is not fixed.  In the same way as
in the local Metropolis algorithm, $s$ is a random number distributed
according to some a priori probability density.
If we choose $s$ to be
uniformly distributed on the interval $[-\varepsilon,\varepsilon]$, the
integrated acceptance rate $P_{acc}$ (as customarily measured in Monte
Carlo simulations) is obtained by averaging $\Omega(s)$ as follows:
\be \label{integrated_acc}
P_{acc}(\varepsilon)=\frac{1}{2\varepsilon}
                 \int_{-\varepsilon}^{\varepsilon} ds\,\Omega(s)  \ \ .
\ee
It turns out to be a good good rule to adjust
the maximum Metropolis step size $\varepsilon$ such that
$P_{acc}(\varepsilon) \approx  50 \%$.

We consider every algorithm that updates stochastic
variables on a hierarchy of length scales as multigrid Monte Carlo
algorithm.  However, there are two different classes of multigrid
algorithms:  multigrid algorithms in a unigrid implementation
and ``explicit'' multigrid algorithms.

In the unigrid formulation one considers nonlocal updates of the
form (\ref{nonlocal}). Updates on the various layers
of the multigrid are formulated on the level of the finest
lattice $\Lambda_0$.
There is no explicit reference to block spin variables $\Phi$ defined on
coarser layers $\Lambda_k$ with $k > 0$.
In addition, unigrid also refers to a
computational scheme: Nonlocal updates are performed directly on
the level of the finest grid $\Lambda_0$ in practical simulations.

In contrast, the explicit multigrid formulation consists of
explicitly calculating conditional Hamiltonians depending on the block
spin variables $\Phi$ on coarser layers $\Lambda_k$.  This formulation
is possible if the conditional Hamiltonians are of the same type or
similar to the fundamental Hamiltonian.  Then, the conditional
probabilities used for the updating on the $k$-th layer can be computed
without always going back to the finest level $\Lambda_0$. Therefore, an
explicit multigrid implementation reduces the computational work on
the coarser layers (see the work estimates below).
At least in free field theory, an explicit multigrid implementation
is possible using $9$-point prolongation kernels in two dimensions
and generalizations thereof in higher dimensions \cite{brandt,sokalrev}.
Generally, an explicit multigrid implementation for interacting
models is only feasable in special cases with piecewise constant
kernels.

An algorithm formulated in the explicit multigrid style can
always be translated to the unigrid language (that is how we are going
to use the unigrid formulation). The reverse is not true, since not all
nonlocal changes of the fundamental field configuration can be
interpreted as updates of a single block spin variable of an explicit
multigrid.
As an example, one can use overlapping
blocks in the unigrid style
by translating the fields by a randomly chosen distance
\cite{hmsun}.

If we formulate our kinematical analysis in the unigrid
language we nevertheless can include all algorithms formulated in the
explicit multigrid style.

The sequence of sweeps through the different layers $\Lambda_k$ of the
multigrid is organized in a periodic scheme called cycle
\cite{koelnporz}.  The simplest scheme is the V-cycle:  The sequence of
layers visited in turn is $\Lambda_0,\Lambda_1,\ldots \Lambda_K,
\Lambda_{K-1}\ldots \Lambda_1$.  More general cycles are characterized
by the cycle control parameter $\gamma$.  The rule is that from an
intermediate layer $\Lambda_k$ one proceeds $\gamma$ times to the next
coarser layer $\Lambda_{k+1}$ before going back to the next finer layer
$\Lambda_{k-1}$.  A cycle control parameter $\gamma > 1$ samples coarser
layers more often than finer layers.  With $\gamma = 1$ we obtain the
V-cycle.  $\gamma = 2$ yields the W-cycle that is frequently used with
piecewise constant kernels.

The computational work estimates for the different cycles are as follows
\cite{sokalrev}: The work for an explicit multigrid cycle is
$\sim L^d$
if $\gamma < l^d$, where $L$ denotes the lattice size.
The work for a unigrid cycle is $\sim L^d \log L$ if $\gamma=1$,
and $\sim L^{d +\log_l \gamma}$ if $\gamma > 1$. Here, $l$ denotes the
blocking factor used in the iterative definition of the block lattices.

If one wants the computational work in the unigrid style
to not exceed (up to a logarithm) an amount
proportional to the volume $L^d$ of the lattice,
one has to use a V-cycle.
Simulations with $\gamma > 1$ (e.g.\ a W-cycle)
can only be performed in the explicit multigrid style.

\section{An approximation formula for $\Omega(s)$}
\label{SECapprox}

In this section we shall derive an approximate formula
for the quantity $\Omega(s)$ defined in (\ref{omega}).
We can write $\Omega(s)$ as
\be\label{intrep}
\Omega(s)= \int du \, \min(1, \mye^{-u}) \int \frac{dp}{2\pi}
\, \mye^{-ipu}
\, \bigl< \mye^{ip \Delta \calH} \bigr> \, .
\ee
Let us assume that the probability distribution of $\Delta \calH$
is approximately Gaussian. We parameterize this
distribution as follows:
\be\label{itsgaussian}
\mbox{dprob}(\Delta\calH) \propto
d \Delta\calH \,
\exp \bigl(- \frac{1}{4h_2} (\Delta\calH - h_1)^2 \bigr) \, ,
\ee
with $h_1 = \EW{\Delta\calH}$ and $h_2 = \half \bigl(
\EW{\Delta\calH^2} - \EW{\Delta\calH}^2 \bigr)$.
Then
\be
\EW{  \exp(ip \delH)} \approx
\exp(ih_1 p - h_2 p^2 )\, .
\ee
The integrations in eq.\ (\ref{intrep}) can be performed
exactly since there are only Gaussian integrals involved. The
result is
\be
\Omega(s) \approx
\half \left( \erfc \biggl(\frac{h_1}{2\sqrt{h_2}}\biggr)
+ \exp(h_2-h_1) \,
\erfc\biggl(\frac{2 h_2-h_1}{2\sqrt{h_2}}\biggr) \right) \, ,
\ee
\noindent
with
$
\erfc(x) = {2/\sqrt{\pi}} \int_x^{\infty} \!
dt \, \exp(-t^2)
$.
We shall now exploit the translational invariance of the
measure
${\cal D} \phi = {\cal D}(\phi+s\psi)$ to show that
the difference of $h_1$ and $h_2$ is
of order $s^4$. The starting point is the observation
that $\EW{\exp(-\delH)} = 1 $. This implies that
\be
\left.
\frac{\partial^n}{\partial s^n}
\ln \EW { \exp(-\delH) } \right\vert_{s=0}
= \left.
\frac{\partial^n}{\partial s^n}
\sum_{m=1}^{n} \frac{1}{m!} \EW{ (-\delH)^m}_c \,
\right\vert_{s=0} = 0  \, .
\ee
$\EW{(.)}_c$ denotes the connected (truncated) expectation value.
Note that there are no contributions in the sum
for $m > n$. This follows from the fact that
$\delH$ is of order $s$. Consequently,
$(\delH)^m = O(s^m)$, and all contributions with $m > n$ vanish
in the limit $s \rightarrow 0$.
For $n$=2 we obtain the relation
\be\label{SD}
\left.
\frac{\partial^2}{\partial s^2}
\left( \EW{-\delH} + \half \EW{\delH^2}_c \right)\, \right\vert_{s=0} =
\left.
\frac{\partial^2}{\partial s^2}
( - h_1 + h_2 )
\right\vert_{s=0} = 0 \, .
\ee
If we assume that $h_1$ and $h_2$ are even in $s$ (which is the
case if $\calH$ is even in $\phi$), then eq.\ (\ref{SD}) says
that the difference of $h_1$ and $h_2$ is of order $s^4$.

We shall later demonstrate that the approximation
$h_1 \approx h_2$ is in practice very good, even for small
block size.
In this case the acceptance rate prediction simplifies further,
\be\label{formula}
\Omega(s) \approx
\erfc ( \half \sqrt{h_1} ) \, .
\ee
(For an analog result in the context of
hybrid Monte Carlo see \cite{hybrid_cumul}.)

For free massless field theory with Hamiltonian
$\calH(\ph)= \half(\ph,-\Delta \ph)$, we get
$h_1 = h_2=\half\alpha\, s^2$ with $\alpha =(\psi,-\Delta\psi)$,
and our approximation formula becomes {\em exact}:
\be\label{omega_free}
\Omega(s) = \erfc \left(
\sqrt{\frac{\alpha}{8}}
 \, \vert s \vert \right) \, .
\ee
Eq. (\ref{omega_free}) can be checked directly by using
$\EW{\exp(ip \delH)} = \exp(ih_1 p - h_2 p^2 )$ in eq.\ (\ref{intrep}).
This relation is exact in free field theory.

\section{Coarse-to-fine interpolation}

\label{SECkernels}

In this section we shall discuss several choices of the coarse-to-fine
interpolation kernels. In order to have a ``fair'' comparison,
all kernels $\psi$ will be normalized according to eq.\ (\ref{normpsi}).

In free massless field theory, the quantity $\alpha = (\psi,-\Delta\psi)$
characterizes the decrease of the acceptance rate $\Omega(s)$ eq.\
(\ref{omega_free}) with increasing shift $s$.  Therefore it is natural
to minimize $\alpha$ in order to maximize $\Omega(s)$ for fixed $s$.

The optimal kernel $\psi^{exact}$ from the point of view of acceptance
rates can be defined as follows: minimize the quadratic form
\be\label{qform}
\alpha\,=\,(\psi,-\Delta \psi)
\ee
under the constraints that the average of $\psi$ over the
``central block'' $x_o'$ is given by $L_B^{(2-d)/2}$,
and its average over blocks $x' \neq x_o'$ vanishes:
\be
L^{-d}_B \sum_{x \in x'} \psi_x = L^{(2-d)/2}_B \delta_{x',x_o'}
\, \mbox{\ for all \ } x' \in \Lambda_k    \ \ .
\ee
This variational problem can be solved with the help of
Fourier methods. The result is
\be
\psi^{exact}_x = L_B^{\frac{2+d}{2}} {\cal A}_{x,x_o'}\; ,
\ee
where ${\cal A}_{x,x_o'}$ denotes the Gaw\c{e}dzki-Kupiainen kernel
(see, e.g.\, \cite{cargmack}).  The use of this kernel leads to a
complete decoupling of the different layers of the multigrid.  This way
of interpolating from a coarser block lattice $\Lambda_k$ to the fine
lattice $\Lambda_0$ is well known in rigorous renormalization group
theory \cite{rigor}.  It is interesting that considerations about
optimizing acceptance rates in a stochastic multigrid procedure lead to
the same choice of the interpolation kernel.

Because $\psi^{exact}$ is nonvanishing on the whole lattice, it
is not convenient for numerical purposes.  For an attempt to change the
block spin $\Phi_{x_o'}$ on block $x_o'$ one has contributions to the
change of the Hamiltonian from all lattice points.  Therefore the
computational work for a single update is proportional to the volume.

We define a ``truncated kernel'' $\psi^{trunc}$ by restricting the
support of $\psi$ on the block $x_o'$ and its nearest neighbor blocks
$y_o'$
\be
\psi_x^{trunc} = 0 \mbox{\ if \ } x \not \in x_o' \mbox{\ or \ } x
\not \in y_o',
\mbox{\ where \ } y_o'\, n.n. \, x_o'\;.
\ee
In other words, the Laplacian in eq.\ (\ref{qform}) is replaced by a
Laplacian $\Delta_D$ with Dirichlet boundary conditions on the boundary
of the support of $\psi$.  We again minimize $\alpha=(\psi,-\Delta_D
\psi)$ under the $2d+1$ constraints that the average of $\psi$ over the
blocks $x_o'$ and its nearest neighbor blocks is given.  This
minimization can be performed numerically by a relaxation procedure.  In
order to maintain the normalization condition, one always updates
simultaneously two spins residing in the same block, keeping their sum
fixed.  The $\psi^{trunc}$-kernels were used in a multigrid simulation
of the $\phi^4$ model in four dimensions \cite{phifour}.

{}From a practical point of view, it is convenient to use kernels that
have support on a single block $x_o'$, i.e.\
\be
\psi_x = 0 \,\,\,\mbox{if}\;x \not \in x_o'\;.
\ee
We define a kernel $\psi^{min}$ with this property by minimizing
$\alpha=(\psi,-\Delta_{D,x_o'} \psi)$ under the constraint that the
average of $\psi$ over the block $x_o'$ is given.  The Laplacian with
Dirichlet boundary conditions on the boundary of $x_o'$ is defined as
follows:
 \be
  (\Delta_{D,x_o'} \phi )_x =
   \left[ -2d\,\phi_x + \strr \right.
 \sum_{\stackrel{{\scriptstyle y \, n.n. x}}{y \in
x_o'}}
   \left. \strr \phi_{y} \, \right]
    \mbox{\ \ \ for} \;x \in x_o'\ .
 \ee
$\psi^{min}$ can be calculated using an orthonormal set of
eigenfunctions of $\Delta_{D,x_o^{\prime}}$.

We shall now discuss other kernels with support on the block
that are frequently used in the literature.

\vskip5mm \noindent
{\em piecewise constant interpolation:}
\be
 \psi^{const}_x=\left\{
\begin{array}{ll}
{L_B}^{(2-d)/2} &\mbox{for}\; x \in x_o' \\
0&\mbox{for}\; x \not \in x_o' \; .\\
\end{array}
\right.
\ee
This kernel has the advantage that for many models the conditional
Hamiltonians used for updating on coarse grids are of the same type or
similar to the fundamental Hamiltonian.  This means that the
conditional probabilities used for the updating on the $k$-th layer can
be computed without always going back to the finest level $\Lambda_o$.
Therefore, an explicit multigrid implementation with a W-cycle can be
used.

\vskip5mm \noindent
{\em piecewise linear interpolation:} \\
We consider the block
\be
 x_o' = \left\{ x \, \vert \,
 x^{\mu}\ \in \bigl\{1,2,3,\ldots,L_B \bigr\},\,\mu=1,\dots, d \right\}
\, .
\ee
The kernels for other blocks are simply obtained by translation.
For $L_B$ even, $\psi^{linear}$ is given by
\be
 \psi^{linear}_x={\cal N}\prod_{\mu = 1}^{d}
 \left\{\frac{L_B+1}{2}-\left|x^{\mu}-\frac{L_B+1}{2}\right|\right\}
\;\;\;\mbox{for}\; x \in x_o'  \ \ .
\ee
${\cal N}$ is a normalization constant.

\vskip5mm \noindent
{\em ground state projection kernels:} \\
$\psi^{sine}$ is the eigenfunction corresponding to the
lowest eigenvalue of the negative Laplacian with Dirichlet
boundary conditions
$-\Delta_{D,x_o'}$ :
\be
 \psi^{sine}_x = \left\{
\begin{array}{ll}
{\displaystyle {\cal N}\prod_{\mu=1}^d
\sin(\frac{\pi}{ L_B+1 } x^{\mu})}
&\mbox{for}\; x \in x_o' \\
&\\
0&\mbox{for}\; x \not \in x_o' \ \ .\\
\end{array}\right.
\ee
Again, ${\cal N}$ denotes a normalization constant.
Note that this kernel is different from $\psi^{min}$.
A generalization of this kernel was introduced for scalar fields in the
background of nonabelian gauge fields in ref.\  \cite{thomask}.

\begin{table}
 \centering
 \caption[dummy]{\label{tab2} Results for $\alpha=(\Psi,-\Delta\Psi)$
   in 2 dimensions, $512^2$ lattice}
 \vspace{2ex}
\begin{tabular}{|c||c|c|c|c|c|c|c|c|}
\hline\str
 kernel & $L_B$=2   & $L_B$=4   & $L_B$=8  & $L_B$=16
        & $L_B$=32  & $L_B$=64  & $L_B$=128 & $L_B$=256
\\[.5ex] \hline \hline \str
 exact  & 6.899 & 8.902 & 9.705 & 9.941 & 10.00 & 10.02 & 10.18 & 13.11
\tabhline
 trunc  & 7.000 & 9.405 & 10.73 & 11.38 & 11.69 & 11.84 & 11.92 & --
\tabhline
 min    & 8.000 & 13.24 & 18.48 & 22.58 & 25.23 & 26.76 & 27.59 & 28.02
\tabhline
 sine   & 8.000 & 13.62 & 19.34 & 23.78 & 26.62 & 28.25 & 29.13 & 29.58
\tabhline
 linear & 8.000 & 15.80 & 24.58 & 31.84 & 36.68 & 39.51 & 41.05 & 41.84
\tabhline
 const  & 8.000 & 16.00 & 32.00 & 64.00 & 128.0 & 256.0 & 512.0 & 1024
\\[.3ex] \hline
\end{tabular} \end{table}

The results for the quantities $\alpha = (\psi,-\Delta\psi)$ for
different kernels in two dimensions are presented in table \ref{tab2}.
We used a $512^2$ lattice ($\psi^{exact}$ depends on the lattice
size).  The different kernels are ordered according to increasing
value of $\alpha$.

The values of $\alpha^{exact}$ and $\alpha^{trunc}$ are close together.
This shows that the truncation of the support of $\psi$ to the block and
its nearest neighbor blocks is a good approximation
to $\psi^{exact}$ (in the sense of acceptance rates).
The
value of $\alpha^{exact}$ for $L_B=256$ is remarkably higher than on
smaller blocks.  This is a finite size effect because the block lattice
consists only of $2^2$ points.  Since the nearest neighbors overlap on a
$2^2$ lattice, no result for $\alpha^{trunc}$ is quoted for $L_B=256$.
The values of $\alpha$ for the smooth kernels with support on the block
$\psi^{min}, \psi^{sine}$ and $\psi^{linear}$ are of the same magnitude.
We can see that $\psi^{sine}$ is almost as good as the optimal $\psi =
\psi^{min}$.

\begin{table}
 \centering
 \caption[dummy]{\label{tab1} Results for $\alpha=(\Psi,-\Delta\Psi)$
   in 4 dimensions, $64^4$ lattice}
 \vspace{2ex}
\begin{tabular}{|c||c|c|c|c|c|}
\hline\str
 kernel & $L_B$=2 & $L_B$=4 & $L_B$=8 & $L_B$=16 & $L_B$=32
\\[.5ex] \hline \hline \str
 exact  & 14.48 & 20.38 & 23.48 & 24.71 & 30.62
\tabhline
 trunc  & 14.67 & 21.61 & 26.54 & 29.26 & --
\tabhline
 min    & 16.00 & 27.72 & 41.56 & 54.18 & 63.33
\tabhline
 sine   & 16.00 & 30.37 & 48.46 & 64.85 & 76.44
\tabhline
 linear & 16.00 & 39.02 & 70.78 & 101.0 & 122.9
\tabhline
 const  & 16.00 & 32.00 & 64.00 & 128.0 & 256.0
\\[.3ex] \hline
\end{tabular} \end{table}

The results for different kernels in four dimensions are presented in
table \ref{tab1}.  Here we used a $64^4$ lattice.  In principle, the
$\alpha$'s behave as in two dimensions.  The values of $\alpha^{linear}$
for small blocks are higher than $\alpha^{const}$.  The pyramids of the
piecewise linear kernels have a lot of edges in four dimensions which
lead to high costs in the kinetic energy.

The $L_B$-dependence of the $\alpha$'s in $d$ dimensions is
\be
\begin{array}{lcll}
\alpha &=& 2 d L_B \quad &\mbox{for piecewise constant kernels} \, ,
\nonumber \\
\alpha &\rellow{\longrightarrow}{L_B >\!\!> 1}&
\mbox{const} \quad &\mbox{for smooth kernels} \, .
\end{array}
\ee
As an example, the expression for $\alpha^{sine}$ in $d$ dimensions
is
\be
\alpha^{sine}=L_B^{2+d}(L_B+1)^d2^{d+2}d
\sin^{4d+2}\left[\frac{\pi}{2(L_B+1)}\right]
\sin^{-2d}\left(\frac{\pi}{L_B+1}\right) \ \ .
\ee
For large block sizes we find
\be
\alpha^{sine}\rellow{\longrightarrow}{L_B >\!\!> 1}
d\,\frac{\pi^{2d+2}}{2^{3d}}\,=\,const  \ \ .
\ee
{}From table \ref{tab2} we observe that in two dimensions $\alpha$
becomes
almost independent of $L_B$ for the smooth kernels if the block size is
larger than 16.  In four dimensions (table \ref{tab1}), we find
$\alpha(L_B) \sim const$ only for $\alpha^{exact}$.  The other
$\alpha$'s for the smooth kernels have not become independent of $L_B$
for the block sizes studied.

\section{Acceptance rates in free field theory}

\label{SECfree}

Recall that we have $\Omega(s) = \erfc \left(\sqrt{\alpha/8} \, \vert s
 \vert \right)$ in massless free field theory.
$\Omega(s)$ is only a function of the product $\alpha s^2$.
In order to keep $\Omega(s)$ fixed when the block size $L_B$
increases we have to rescale the changes $s$ like $\alpha(L_B)^{-1/2}$.
As a consequence, to maintain a constant
acceptance rate in massless free field theory,
$s$ has to be scaled
down like $L^{-1/2}_B$ for piecewise constant kernels, whereas for
smooth kernels the acceptance rates for large $L_B$
do not depend on the block size.

Note that this behavior of the acceptance rates for large $L_B$ is not
yet reached in four dimensions for the block sizes studied (except for
$\psi^{exact}$).  See also the discussion of the
Metropolis step size below.
At least for free field theory, the disadvantage of the piecewise
constant kernels can be compensated for by using a W-cycle instead of a
V-cycle.  Smooth kernels can be used only in V-cycle algorithms.
An exception are $9$-point prolongation kernels in
two dimensions and generalizations thereof in higher dimensions.
They can also be used
with a W-cycle, at least in free field theory
(cf.\ section \ref{SECunigrid}).

We now illustrate what this rescaling of $s$ means for the
Metropolis step size $\varepsilon$ in an actual multigrid Monte Carlo
simulation.  Look at the integrated acceptance probability defined in
eq.\ (\ref{integrated_acc}).  If we insert the exact result
(\ref{omega_free}) for massless free field theory
we get
\be
P_{acc}(\varepsilon)
=\mbox{erfc}\left(\sqrt{\frac{\alpha}{8}}\varepsilon\right)
+\frac{1}{\sqrt{\frac{\pi\alpha}{8}}\varepsilon}
\left[\str 1-e^{-\frac{\alpha}{8}\varepsilon^2 }\right]\ \ .
\ee
$P_{acc}$ is only a function of the product $\alpha
\varepsilon^2$.  In order to keep $P_{acc}$ fixed (to, e.g.\ 50
percent) we
have to rescale $\varepsilon(L_B)$ like $\alpha(L_B)^{-1/2}$, exactly in
the same way as we had to rescale $s$ to keep $\Omega(s)$ fixed.
This $L_B$-dependence is plotted in figure 1 for two dimensions
and in figure 2 for four dimensions.

We now discuss massive free field theory with Hamiltonian
$\calH(\ph)= \half(\ph,[-\Delta + m^2] \ph)$. We find
$h_1 =\half\alpha_m\, s^2 $, with $\alpha_m$ given by
\be
\alpha_m = (\psi,[-\Delta + m^2] \psi)
         = \alpha
+  m^2 \sum_{x \in \Lambda_0} \psi_x^2 \, .
\ee
Therefore the exact result is
$\Omega(s) = \erfc \left( \sqrt{\alpha_m/8}\, \vert s \vert \right)$.
A term $\sum_x \psi_x^2$ scales $\sim L_B^2 $ in arbitrary
dimensions.
For piecewise constant kernels
\be
\sum_{x \in \Lambda_0}
  \left(\psi^{const}_x \right)^2 \,
                                    = \, L_B^2 \ .
\ee
For $\psi^{sine}$-kernels  we find
\be
\sum_{x \in \Lambda_0}
                       \left(\psi^{sine}_x \right)^2
= 2^{d} L_B^{2+d}(L_B+1)^d
\sin^{4d}\left[\frac{\pi}{2(L_B+1)}\right]
\sin^{-2d}\left(\frac{\pi}{L_B+1}\right) \ \ ,
\ee
and for large block sizes
\be
\sum_{x \in \Lambda_0}\left(\psi^{sine}_x \right)^2
\rellow{\longrightarrow}{L_B >\!\!> 1}
\frac{\pi^{2d}}{2^{3d}} L_B^2 \ \ .
\ee
If the block size $L_B$ is smaller than the correlation
length $\xi = 1/m$, $h_1$ is still dominated by the kinetic term
$s^2 (\psi,-\Delta\psi)$,
and the discussion is the same as in the massless case.

As soon as the block size $L_B$ becomes larger than $\xi$, $h_1$ is
dominated by the mass term $s^2 m^2 \sum_x \psi^2_x \sim s^2 L_B^2$, and
$s$ has to be rescaled like $s \sim L_B^{-1}$ in order to maintain
constant acceptance rates.  Of course this is a dramatic decrease for
large block sizes compared to $s \sim const$ (using smooth kernels) in
the massless case.  Block spins on large blocks are essentially
``frozen''.  But this is not dangerous for the performance of the
algorithm in massive free field theory: The effective probability
distribution for the block spins $\Phi$ is given by
$\exp(-\calH_{\mbox{\scriptsize eff}}(\Phi))$, where
$\calH_{\mbox{\scriptsize eff}}(\Phi)$ denotes the
effective Hamiltonian in the sense of the block spin renormalization
group \cite{schladming}.  The physical fluctuations of the block spins
are dictated by an effective mass term
\be
 m_{\mbox{\scriptsize eff}}^2 \sum_{x' \in \Lambda_k} \Phi_{x'}^2\,
\; \;\; \mbox{with}\;\;\; m_{\mbox{\scriptsize eff}}^2 \sim m^2
                             L_B^2  \, .
\ee
Thus, the algorithmic fluctuations (described by the mass term $m^2
\sum_x \psi_x^2 \sim m^2 L_B^2$) and the physical fluctuations
(described by the effective mass $\sim m^2 L_B^2$) behave similar, and
the multigrid algorithm is able to create fluctuations just of the size
that is needed by the physics of the model.
Moreover there is no need to do updates at length scales
larger than $\xi$ in order to beat CSD.

In this sense, the discussed algorithmic mass term $m^2 \sum_x \psi^2_x$
is well behaved for free field theory, since it decreases with the
physical mass in the vicinity of the critical point.  As we shall see in
section \ref{SECappl}, for interacting models close to criticality, a
different scenario is possible.  There, it can happen that an
algorithmic ``mass term'' $\sim \sum_x \psi_x^2$ persists, whereas the
renormalized mass vanishes.  If this happens, the multigrid algorithm is
not able to produce the large critical fluctuations required by the
physics, and we can {\em not} expect that CSD will be eliminated.

The $L_B$-dependence of a term $\sum_x \psi_x^4$ will also be
needed in the study of the $\phi^4$ theory in section \ref{SECappl}
below.
In $d$ dimensions such a term scales $\sim L_B^{4-d}$:
For piecewise constant kernels
\be
\sum_{x \in \Lambda_0}
                       \left(\psi^{const}_x \right)^4
                       \, = \, L_B^{4-d} \ ,
\ee
whereas using $\psi^{sine}$-kernels  we find
\be
\sum_{x \in \Lambda_0}
                       \left(\psi^{const}_x \right)^4
=6^d L_B^{4+2d}(L_B+1)^{d}
\sin^{8d}\left[\frac{\pi}{2(L_B+1)}\right]
\sin^{-4d}\left(\frac{\pi}{L_B+1}\right) \ \ .
\ee
In the limit of large block sizes this term behaves like
\be
\sum_{x \in \Lambda_0}
                       \left(\psi^{sine}_x \right)^4
\rellow{\longrightarrow}{L_B >\!\!> 1}
\left(\frac{3\pi^4}{128}\right)^d L_B^{4-d} \ \ .
\ee

In order to summarize the different large-$L_B$-behavior of local
operators in the kernel $\psi$ discussed here, let us introduce the
{\em degree of relevance} in the sense of the perturbative
renormalization group:
The (superficial) degree $r$ of relevance of a local operator in
$\psi$ which is a polynomial of $m$ scalar fields
with $n$ derivatives is defined by $r=d+m(2-d)/2-n$.
This definition is valid for smooth kernels.
For large $L_B$, an operator with degree of relevance
$r$ behaves like $L_B^r$.
An operator is called relevant if $r > 0$.
As we have seen in the examples above, a mass term has $r=2$, and a
$\psi^4$-term has $r=4-d$.  A kinetic term $\alpha=(\psi,-\Delta \psi)$
has $r=0$ for smooth kernels.

The only difference for piecewise constant kernels is that a kinetic
term behaves like $\alpha=(\psi,-\Delta \psi) \propto L_B$.

\section{Acceptance rates for interacting models}

\label{SECappl}

In this section, we shall apply formula (\ref{formula}) in the discussion
of multigrid procedures for three different spin models in two
dimensions:  the Sine Gordon model, the XY model, and the
single-component $\ph^4$ theory.  The scale dependence of acceptance
rates for interacting models will be compared with the behavior in
free field theory, where CSD is known to be eliminated
by a multigrid algorithm.

\subsection{2-dimensional Sine Gordon model}

\label{SUBSECsinegordon}

The 2-dimensional Sine Gordon model is defined by the
Hamiltonian
\be
\calH(\ph) = \frac1{2\beta} ( \ph, -\Delta \ph)
- \zeta \sum_x \cos \ph_x  \, .
\ee

The model undergoes a Kosterlitz-Thouless phase transition at $\beta_c$.
In the limit of vanishing fugacity $\zeta$, the location of the critical
$\beta$ is exactly known:  $\beta_c \rightarrow 8 \pi$ for $\zeta
\rightarrow 0$.  For $\beta > \beta_c$, the model is in the massless
phase, and the flow of the effective Hamiltonian (in the sense of the
block spin renormalization group) converges to that of a massless free
field theory:  the long distance behavior of the theory is that of a
Gaussian model.  Since multigrid algorithms have proven to be efficient
in generating long wavelength Gaussian modes, one might naively conclude
that multigrid should be the right method to fight CSD
in the simulation of the Sine Gordon model in the massless phase.  But
this is not so.  For $h_1$ we find the expression
\be\label{h1h2}
h_1
= \frac{\alpha}{2\beta} s^2 + \zeta C \sum_x \lbrack 1 - \cos(s\psi_x)
\rbrack \, , \ee
with $C = \langle \cos \ph_x \rangle$.
Recall that $h_1$ is the quantity that determines the
acceptance rates $\Omega(s)$:
\be
\Omega(s) \approx \erfc ( \half \sqrt{h_1} ) \, .
\ee
The essential point is that the second term in (\ref{h1h2}) is
proportional to the block volume $L^2_B$ for piecewise constant
{\em and} for smooth kernels (cf.\ the discussion in section
\ref{SECfree}).
This can be checked for small $s$ by expanding in $s$.
One therefore has to face a dramatic decrease of acceptance when the
blocks become large, even for small fugacity $\zeta$.  A constant
acceptance rate is achieved only when the proposed steps are scaled down
like $L^{-1}_B$.  It is therefore unlikely that any multigrid algorithm
- based on nonlocal updates of the type discussed in this paper - will
be successful for this model.

We demonstrate the validity of formula (\ref{formula}) (using a Monte
Carlo estimate for $C$) by comparing with Monte Carlo results at $\beta
= 39.478$ and $\zeta=1$.  This point is in the massless phase,
where the correlation length $\xi$ is or the order of the lattice
size $L$.  In
figure 3 we show both the numerical and analytical results for
$\Omega(s)$ for $L_B=4,8,16,32$ on lattices of size
$16^2,32^2,64^2,128^2$, respectively.

We tested the precision of our approximation formula for piecewise
constant kernels only.  However, we have no doubts that the quality of
the approximation is also very good for other shape functions $\psi$.

\subsection{2-dimensional XY model}
\label{SUBSECxy}

We now discuss the
2-dimensional XY model, defined by the partition function
\be \label{xy_part}
Z = \int \prod_x d\Theta_x \, \exp \bigl(
\beta \sum_{<x,y>} \cos(\Theta_x-\Theta_y) \bigr) \, .
\ee
The sum is over all unordered pairs of nearest  neighbors in the
lattice.
As the Sine Gordon model, the XY model has a massless (spin wave) phase
for $\beta > \beta_c$, and a massive phase for $\beta < \beta_c$.  The
best available estimate for the critical coupling is $\beta_c =
1.1197(5)$ \cite{matching}.

Nonlocal updates are defined by
\be \label{xy_update}
\Theta_x \rightarrow
\Theta_x + s \psi_x \, ,
\ee
with $\psi$ obeying again the normalization
condition (\ref{normpsi}). To define a (linear) block spin, we
rewrite the partition function (\ref{xy_part}) in terms of
2-component unit vector spin variables $s_x$:
\be
Z = \int \prod_x \left( d^2 s_x \,  \delta(s_x^2-1) \right)
\exp \bigl( \beta \sum_{<x,y>} s_x \cdot s_y \bigr)
\ee
The block spins $S_{x'}$ are then defined as block averages
of the unit vectors $s_x$.

Note that the proposal (\ref{xy_update}) changes the block
spin by an amount $\approx s$ only when the spins inside the
block are sufficiently aligned. This will be the case in the spin wave
phase for large enough $\beta$. For smaller $\beta$, the
correct (or ``fair'') normalization of the kernels $\psi$ is
a subtle point. We believe, however, that our argument is
not affected by this in a qualitative way.

The relevant quantity $h_1$ is given by
\be
h_1 = \beta E \sum_{<x,y>}
\lbrack 1 - \cos( s( \psi_x - \psi_y) )\rbrack \, ,
\ee
with
$E=\langle \cos(\Theta_x - \Theta_y) \rangle $,
$x$ and $y$ nearest neighbors.
For piecewise constant kernels,
$h_1$ is proportional to $L_B$.
For smooth kernels $h_1$ will become independent of $L_B$ for large
enough blocks. For small $s$,
\be
h_1 \approx \half s^2 \beta E \sum_{<x,y>} (\psi_x -\psi_y)^2
= \half s^2 \beta E \alpha \, .
\ee
As above, $\alpha=(\psi,-\Delta \psi)$. This quantity becomes nearly
independent of $L_B$ already for $L_B$ larger than $16$
(cf.\ section \ref{SECkernels}).

{}From the point of view of acceptance rates the XY model therefore
behaves like massless free field theory.  A dynamical critical
exponent $z$ consistent with zero was observed in the massless phase
\cite{xy}.  The failure of multigrid Monte Carlo in the massive phase
($z \approx 1.4$ for piecewise constant kernels \cite{xy}) is an example
for the fact that good acceptance rates are not sufficient to overcome
CSD.

We again checked the accuracy of formula (\ref{formula}) by comparing
with Monte Carlo results at $\beta=1.2$ (which is in the massless
phase, where the correlation length $\xi$ is of the order of the lattice
size $L$).
The only numerical input for the analytical formula was the
link expectation value $E$.  The results are displayed in figure 4.

One can do a similar discussion for the $O(N)$ nonlinear $\sigma$-model
with $N>2$, leading to the same prediction for the scale dependence of
the acceptance rates.  This behavior was already observed in multigrid
Monte Carlo simulations of the $O(3)$ nonlinear
$\sigma$-model in two dimensions
               with smooth and piecewise constant kernels \cite{hmm}.

\subsection{2-dimensional $\phi^4$ theory}
\label{SUBSECphi4}

Let us now turn to single-component
$d$-dimensional $\ph^4$ theory, defined
by the Hamiltonian
\be
\calH(\ph) = \half(\ph, -\Delta \ph)
           + \frac{m_o^2}2 \sum_x \ph_x^2
           + \frac{\lambda_o}{4!} \sum_x \ph_x^4 \, .
\ee

For $h_1$ one finds
\be
h_1 = s^2 \, \left\{
\half \alpha + \lbrack \frac{m_o^2}{2} + \frac{\lambda_o}{4}
P \rbrack \sum_x  \psi_x^2 \right\}
+ s^4 \, \frac{\lambda_o}{4!} \sum_x \psi_x^4 \, ,
\ee
where $P=\langle \ph_x^2 \rangle$.  We have used that expectation values
of operators which are odd in $\ph$ vanish on finite lattices.
Recall that $\sum_x \psi_x^2$
increases with $L^2_B$, independent of $d$, whereas $\sum_x \psi_x^4$
scales like $L^{4-d}_B$, for smooth and for piecewise constant kernels
(cf.\ the discussion of the different kernels
in section \ref{SECfree}).
We conclude that also in this model we have to face rapidly decreasing
acceptance rates when the blocks become large.  As in the case of the
Sine Gordon model, $s$ has to be rescaled like $L^{-1}_B$ in order to
maintain constant acceptance rates.

Therefore there is little hope that multigrid algorithms of the type
considered here can overcome CSD in the
1-component $\ph^4$ model.  In numerical simulations of 2-dimensional
$\ph^4$ theory, a dynamical critical behavior is found that is
consistent with $z \approx 2$ for piecewise constant and for smooth
kernels \cite{sokalprl,sokalrev,linn}.  In four dimensions, there is no
definite estimate for $z$ \cite{phifour}.

Figure 5 shows a comparison of Monte Carlo results for
2-dimensional $\ph^4$ theory with the theoretical prediction based on
the numerical evaluation of $P$.  The simulations were done in the
symmetric phase at $m_o^2 = - 0.56$ and $\lambda_o = 2.4$.  The
correlation length at this point is $\xi \approx 15$ \cite{linn}.
\subsection{Summary of section 6}

Our approximation formula has proven to be quite precise.
The results for three different models are consistent with the following
rule:

\noindent
{\sl Sufficiently high acceptance rates for a complete elimination
of CSD can only be expected if
$ h_1 = \langle \calH(\ph+s \psi) - \calH(\ph) \rangle $
contains no relevant operator in $\psi$. }

As we have seen above, the typical candidate for a relevant operator
in $\psi$ is always  an ``algorithmic mass term''
of the type $\sim s^2 \sum_x \psi_x^2$ with degree of relevance
$r=2$.

This rule is formulated for smooth kernels.
For piecewise constant kernels, it has to be modified.
There, the kinetic term $\alpha \propto  L_B$ is relevant as well.
At least in free field theory this disadvantage can be compensated for
by using a W-cycle.
Apart from this modification the rule carries over to the case
of piecewise constant kernels.

\section{A multigrid procedure for lattice gauge fields}

\label{SECgauge}

In this section we propose a multigrid procedure for
pure lattice gauge theory and study the
behavior of acceptance rates with increasing block size $L_B$.

We consider partition functions
\be
Z = \int \prod_{x,\mu} d U_{x,\mu} \,
\exp\bigl( - \calH(U) \bigr) \, .
\ee

The link variables $U_{x,\mu}$ take values in the gauge
group $U(1)$ or $SU(N)$, and $dU$ denotes the corresponding
invariant Haar measure.
The standard Wilson action $\calH(U)$ is given by
\be \label{wilson-action}
{\cal H}(U)\,=\,\beta\sum_{\cal P} \bigl[ 1 -
\foN \mbox{Re} \, \Tr \, U_{\cal P} \bigr] \ \ .
\ee
The sum in (\ref{wilson-action}) is over all plaquettes in the
lattice. The $U_{\cal P}$ are path ordered
products around plaquettes ${\cal P}$,
\be
U_{\cal P} \,=\, \Um{x}\Un{x+\hat{\mu}}
                 \Um{x+\hat{\nu}}^{*}\Un{x}^{*}\ \ .
\ee
$U^*$ denotes the hermitean conjugate (= inverse) of $U$.
\subsection{The abelian case}
\label{SECabel}

We now consider the case of gauge group $U(1)$. The link variables
are parameterized with an angle
$ -\pi \leq \theta_{x,\mu} < \pi$ through
\be
\Um{x}\,=\,\exp( i \theta_{x,\mu}) \ \ .
\ee
Nonlocal updates can be defined as follows:
One chooses a hypercubic block $x_o'$ of size $L_B^d$
and a direction $\tau$ with $1 \leq \tau \leq d$.
During the update, $\tau$ will be kept fixed.
All the link variables $\Ut{x}$ attached to sites $x$ inside the block
$x_o'$ are proposed to be changed simultaneously:
\be
\Ut{x} \rightarrow \exp(is\psi_x)\Ut{x} \, .
\ee
Again, $\psi$ denotes an interpolation kernel
as introduced in section \ref{SECkernels}.
This updating scheme
was introduced and studied in two
dimensions in ref.\ \cite{laursen} and also in four dimensions
in ref.\  \cite{newgauge}, using piecewise constant kernels.
Of course, one can use all
versions of smooth kernels as well, with their support not necessarily
restricted to the block $x_o'$.

Let us now study the acceptance rates for these update proposals
with the help of formula (\ref{formula}). We consider
general kernels $\psi$.
For $h_1 = \langle \Delta \calH \rangle$ we find
\be \label{gaugeu1}
h_1\,=\,
\beta P \sum_{x \in \Lambda_0}
\sum_{\mu \neq \tau}
\left[ 1-\cos\bigl( s (\psi_{x+\hat{\mu}}-\psi_x)\bigr) \right] \ \ ,
\ee
with $P = \EW{ \Tr U_{\cal P} }$.
The sum does not include contributions from links
which point in the fixed $\tau$-direction. If we denote
the ``slice'' of lattice sites
with ${\tau}$-component $t$ as
$ \Lambda_t^{\tau} =
\left\{ x \in \Lambda_0 \, \vert \, x_{\tau} = t \right\}$,
we see that
$h_1$ is a sum of independent contributions from
slices orthogonal to the $\tau$-direction.
Therefore, no smoothness
of kernels is needed in the $\tau$-direction, and from now on we
choose $\psi_x$ to be constant in that direction.
Let us assume that the support of $\psi$ in $\tau$-direction
is contained in the block $x_o'$. Then we find for small $s$
\be
h_1\,\approx \,
\half s^2 \beta P L_B \sum_{x \in \Lamt}
\sum_{\mu \neq \tau}
( \psi_{x+\hat\mu} - \psi_{x} )^2 =
\half s^2 \beta P \alpha_{d-1} \ \ .
\ee
with $\alpha_{d-1} = (\psi',-\Delta \psi') $.
Here, $\psi'$ denotes the kernel $\psi$ restricted to the
$d-1$ dimensional sublattice $\Lamt$, multiplied with
a factor $L_B^{1/2}$ in order to be properly normalized
as a $(d-1)$ dimensional kernel.


{}From the kinematical point of view, the behavior of acceptance rates
in the $U(1)$ lattice gauge theory in $d$ dimensions is the same as in
massless free field theory.  One might therefore expect that it is
possible for a multigrid algorithm to overcome CSD in
this model.  Indeed, in numerical simulations in two dimensions using
piecewise constant kernels, the dynamical critical exponent was found to
be $z \approx 0.1$ \cite{laursen}.
However, it was also observed that the multigrid algorithm
is not able to move efficiently between different topological
sectors. The above quoted exponent should therefore be
interpreted with some care.
For the results of a study of a multigrid algorithm for
4-dimensional $U(1)$-theory see \cite{newgauge}.

Let us conclude the discussion of the abelian case
with the remark that with no
loss of generality one could consider blocks $x_o'$ that
consist only of one layer in $\tau$-direction, i.e., effectively
$(d\!-\!1)$-dimensional blocks. This is a consequence of the fact
that the updates of the two variables $\Ut{x}$ and $\Ut{y}$
are statistically independent if $x_{\tau} \neq y_{\tau}$.
This property carrys over to the nonabelian case.

\subsection{The nonabelian case: gauge group $SU(2)$}
\label{SUBSECsu2}

\subsubsection{Covariant nonlocal update proposal}
\label{SUBSUBSECcov}

We shall now discuss a generalization of the above described procedure
to the nonabelian case.  To be concrete, we study 4-dimensional $SU(2)$
lattice gauge theory.

Let us start with an attempt of a straightforward generalization
of the procedure described for the abelian theory.
This would amount to propose updates
\be \label{silly}
\Ut{x} \rightarrow \Ut{x}' = R_x \, \Ut{x} \, ,
\ee
where the ``rotation'' matrices $R_x$ are in $SU(2)$.
We parametrize them as
\be
R_x(\vn,s) = \cos( s \psi_x /2 )
 + i \sin( s \psi_x /2) \,
\vn \!\cdot\! \vs \, ,
\ee
where $\vn$ denotes a three-dimensional real unit vector,
and the $\sigma_i$ are Pauli matrices.
$\psi$ will
have support on 3-dimensional blocks $x'$ of size $L_B^3$,
and the blocks consist of points lying entirely in $\Lamt$.

We use the fact that updates of link variables in different slices
are statistically independent (as discussed at the end of subsection
\ref{SECabel}).  One possible updating scheme is to perform the proposed
updates on different slices in sequence.  Another possible updating
scheme consists of building hypercubic four-dimensional blocks out of
''staples'' of $L_B$ three-dimensional blocks of size $L_B^3$ and to
perform the updates on this hypercubic block simultaneously.
Because of the
independence of different slices, the analysis of acceptance
rates is the same for both cases.  For simplicity we study
three-dimensional blocks here.

The energy change associated with the update proposal (\ref{silly}) is
\be
\Delta \calH = - \frac{\beta}2
\sum_{\calP} \Tr \bigl( U_{\calP}' - U_{\calP} \bigr)
= - \frac{\beta}2 \sum_{x \in \Lamt}
\sum_{\mu \neq \tau} \Tr \bigl\{
( R_x^* \Um{x} R_{x+\hat\mu} - \Um{x} ) H_{x,\mu}^* \bigr\} \, ,
\ee
with $H_{x,\mu}^* = \Ut{x+\hat\mu} \Um{x+\hat\tau}^* \Ut{x}^*$.
The relevant quantity for the acceptance rates is
$h_1 = \EW{\Delta \calH}$. For piecewise constant kernels
$\psi$ one gets
\be \label{gauge_deltah}
h_1 = \frac{3 \beta}2 A' \, (L_B -1) L_B^2 \, \sin^2(s L_B^{-1/2}/2) +
\, 3 \beta P \, L_B^2 \bigl[ 1- \cos(s L_B^{-1/2}/2) \bigr] \, ,
\ee
with
\bea
A' &=& -
\bigl\langle \Tr \bigl((\vn \!\cdot\! \vs \, \Um{x} \,
 \vn \!\cdot\! \vs - \Um{x})
H_{x,\mu}^* \bigr) \bigr\rangle \, ,
\nonumber \\
P &=&
\bigl\langle \Tr \bigl( \Um{x}
H_{x,\mu}^*\bigr) \bigr\rangle
= \bigl\langle \Tr U_{\calP} \bigr\rangle \,.
\eea

The first contribution to $A'$ is the expectation value of
a quantity that is not gauge invariant. Determining its
gauge invariant projection, we can show that this contribution
vanishes:
\be
\bigl\langle \Tr \bigl( \vs \cdot \vn \, \Um{x} \, \vs \cdot \vn
H_{x,\mu}^* \bigr) \bigr\rangle
=
\int d V
\bigl\langle \Tr \bigl( \vn \!\cdot\! \vs \, V \Um{x} \, \vn \!\cdot\!\vs
H_{x,\mu}^*  V^* \bigr) \bigr\rangle = 0 \, ,
\ee
because for $SU(2)$
\be\label{trace_formula}
\int d V \Tr(AVBV^*) = \half \Tr A \, \Tr B \, ,
\ee
and the Pauli matrices are traceless.
Therefore we get  $A' = P$.

To the first term in eq.\ (\ref{gauge_deltah}) all links contribute that
are entirely inside the block, whereas the second term sums the
contributions of all links that have one site in common with the surface
of the block.  For small $s$, the first term behaves like $ s^2 L_B^2 $.
This is exactly the behavior of a mass term that, as we have learned in
the previous sections, is toxic for the multigrid algorithm.

The main difference to the abelian case is that in addition to the
costs from the surface of the
block we have a contribution from the interior of the block.
Unfortunately, this contribution grows quadratic with the block
dimension $L_B$.

This does not come as a surprise.  Due to the gauge invariance
of the model, the $U$'s do not have a gauge invariant meaning.
Therefore the rotations $R_x$ that are smooth over the block for a given
gauge can be arbitrarily rough after a gauge
transformation.  It is therefore clear that the rotation matrices have
to be chosen in a gauge covariant way.

We generalize the update proposal (\ref{silly}) as follows:
\be \label{clever}
\Ut{x} \rightarrow \Ut{x}' = g_x^* R_x g_x \, \Ut{x} \, ,
\ee
with $g_x \in SU(2)$.
Note that in the abelian case we obtain nothing new, because
$g_x$ and $R_x$ commute. In the nonabelian case, we find for
piecewise constant $\psi$
\be \label{gauge_deltah_clever}
h_1 = \frac{\beta}{2} \sum_{(x,x+\hat\mu) \in x_o'}
A_{x,\mu} \, \sin^2(s L_B^{-1/2}/2) +
\, 3 \beta P \, L_B^2 \bigl[ 1- \cos(s L_B^{-1/2}/2) \bigr] \, ,
\ee
with
\be \label{axmu}
A_{x,\mu} = - \bigl\langle \Tr \bigl((\vn \!\cdot\! \vs \,
\Um{x}^g \, \vn \!\cdot\! \vs - \Um{x}^g)
H_{x,\mu}^{g\,  *} \bigr) \bigr\rangle \, .
\ee
Here we have introduced the notation
$U^g_{x,\mu} = g_x \Um{x} g_{x+\hat\mu}^*$. $H^g$ is defined
analogously.
$U^g$ is the gauge field obtained by applying a gauge transformation
$g$ to $U$.  We are free to choose $g$.  To obtain a valid Monte Carlo
algorithm, we require that the $g$'s should not depend on the link
variables to be updated, i.e.\ those living on the links $(x,x +
\hat\tau)$.  On the other hand we want to minimize $h_1$.  Let us
inspect the quantity $A_{x,\mu}$ defined in eq.\ (\ref{axmu}) that leads
to the unwanted mass term behavior of $h_1$.  Consider the extreme case
of $\beta \rightarrow \infty$.  Then the allowed configurations are pure
gauges, i.e.\ configurations that are gauge equivalent to $\Um{x} = 1$
for all $x,\mu$.  If we choose $g$ as the transformation that
transforms all links to unity, it is obvious that $A_{x,\mu}$ is zero.
In particular,
to obtain this result, it is sufficient to gauge all links inside the
block that do not point in the $\tau$-direction to unity.  This
consideration leads to following proposal:  Choose $g$ as the gauge
transformation that maximizes the functional
\be\label{coulomb_gauge}
G_{C,x_0'}(U,g) = \sum_{(x,x+\hat\mu) \in x_o'} \Tr \bigl(
g_x \Um{x} g_{x+\hat\mu}^*\bigr) \, .
\ee
We call this gauge ``block Coulomb gauge''.
This gauge will bring the links inside the block as close
to unity as possible thus leading to a kind of minimization
of $A_{x,\mu}$ (corresponding to a minimization of the
mass term). Note however, that we do not intend to actually
perform the gauge transformation. We use the concept of
gauging only to define covariant rotations $g_x R_x g_x^*$.
Covariance here means that the relevant quantity
$
\Tr \bigl((\vn \!\cdot\! \vs \,
\Um{x}^g \, \vn \!\cdot\! \vs - \Um{x}^g)
H_{x,\mu}^{g\,  *} \bigr)
$
is now gauge invariant. To see this, assume that we pass
from $U$ to $U^h$ by applying the gauge transformation $h$.
The Coulomb gauge condition will then lead to a new
$g' = gh^*$. Now note that
$ (U^h)^{gh^*} = U^g$. The same argument applies to $H$.

Let us summarize the steps of the nonlocal updating scheme for
$SU(2)$:
\begin{enumerate}

\item Choose a block $x_o'$ of size $L_B^3$ that is contained
in the slice $\Lamt$. All link variables $\Ut{x}$ pointing from
sites $x$ inside the block in $\tau$-direction will be moved
simultaneously.

\item Find the gauge transformation $g$ defined by the
block Coulomb gauge condition
\be
G_{C,x_0'}(U,g) = \sum_{(x,x+\hat\mu) \in x_o'} \Tr \bigl(
g_x \Um{x} g_{x+\hat\mu}^*\bigr)
 \,\stackrel{\mbox{!}}{=} \, \mbox{maximal}\, .
\ee

\item Propose new link variables $\Ut{x}'$ by
\be
\Ut{x} \rightarrow \Ut{x}' = g_x^* R_x g_x \, \Ut{x} \, ,
\ee
with
\be
R_x(\vn,s) = \cos( s \psi_x /2 )
 + i \sin( s \psi_x /2) \,
\vn \!\cdot\! \vs \, .
\ee
$s$ is a uniformly distributed random number from the
interval $[-\varepsilon,\varepsilon]$, and $\vn$ is a
vector selected randomly from the three dimensional unit sphere.

\item Calculate the associated change of the Hamiltonian  $\Delta \calH$
and accept the proposed link variables with probability $ \min
\lbrack 1,\exp(-\Delta \calH) \rbrack $.
\end{enumerate}

The detailed balance condition is fulfilled
by this updating scheme:  For the naive version with $g = 1$ it is
straightforward to show that the detailed balance condition holds, since
the rotation matrices $R_x$ are chosen according to a probability
distribution which is symmetric around unity.

If we now take $g$ according to some gauge condition, we have to be
careful that we get the same $g$ before and after the move $\Ut{x}
\rightarrow \Ut{x}'$.  Otherwise this move would not be reversible.  In
other words:  The gauge condition yielding $g$ must not depend on
$\Ut{x}$.  This is indeed the case, since only link variables $\Um{x}$
with $\mu \neq \tau$ enter in the block Coulomb gauge functional.  Note
that we do not have to fix the gauge perfectly.  If we always use the
same procedure in finding $g$ (e.g.\ a given number of relaxation sweeps
starting from $g=1$), we will always get the same $g$ and the nonlocal
update is reversible.

As usual we now choose different (possibly overlapping) blocks $x'$,
different block sizes $L_B$, different slices $\Lamt$ and different
orientations $\tau$ of the slices in turn.


\subsubsection{Acceptance analysis for nonlocal $SU(2)$-updates}
\label{SUBSUBSECaca}

First numerical studies revealed that there is no substantial difference
in the acceptance rates when instead of using the block Coulomb gauge
condition one uses the Coulomb gauge condition for the whole slice
$\Lamt$:
\be\label{coulomb_gauge_slice}
G_C(U,g) = \sum_{(x,x+\hat\mu) \in \Lamt} \Tr \bigl(
g_x \Um{x} g_{x+\hat\mu}^*\bigr)
 \,\stackrel{\mbox{!}}{=} \, \mbox{maximal} \, .
\ee
{}From a practical point of view this gauge condition is very
convenient, because the relaxation algorithm to determine
the $g_x$ can then be vectorized in a straightforward way.

If we use the gauge condition (\ref{coulomb_gauge_slice}), the quantity
$A_{x,\mu}$ becomes translation invariant and also independent of $\mu$
(where we still keep $\mu \neq \tau$).  We get
\be \label{h1_constant}
h_1 = \frac{3 \beta}2 A \, (L_B -1) L_B^2 \, \sin^2(s L_B^{-1/2}/2) + \,
3 \beta P \, L_B^2 \bigl[ 1- \cos(s L_B^{-1/2}/2) \bigr] \, , \ee with
\be A= - \bigl\langle \Tr \bigl((\vn \!\cdot\!  \vs \, \Um{x}^g \,
\vn\!\cdot\!\vs - \Um{x}^g) H_{x,\mu}^{g\, *} \bigr) \bigr\rangle
\ee
Following the discussion after eq.\ (\ref{trace_formula}),
we identify
the square root of $A$ with a ``disorder mass'' $m_D$,
\be
m_D = \sqrt{A} \, .
\ee

%
$m_D$ has the dimension of a mass. It
vanishes in the limit $\beta \rightarrow \infty$, just like physical
masses in the theory.
Because of the disorder inside the block, $m_D$ is
nonzero for finite $\beta$.  This would not be a problem if
for large correlation length $m_D$ scaled like a physical mass
(cf.\ the discussion for massive free field theory in
section \ref{SECfree}).

\subsubsection{Monte Carlo study of $m_D$}

We computed $m_D$ by Monte Carlo simulations for several values of
$\beta$.  To maximize $G_C$ we used a standard Gauss-Seidel relaxation
algorithm vectorized over a checkerboard structure.  The relaxation
procedure consists in going through the lattice and
minimizing the gauge functional (\ref{coulomb_gauge_slice}) locally.
%
For production runs it would be advantageous to use an accelerated gauge
fixing algorithm such as overrelaxation or multigrid \cite{gaugefix}.
In the Monte Carlo studies reported in this section, we always used 50
Gau\ss -Seidel sweeps to determine $g$.  Note that by this procedure,
$G_C$ is not entirely maximized, especially on very large lattices where
the relaxation algorithm suffers from CSD.  However,
for the detailed balance to be fulfilled, we only need that one uses
always the same number of relaxation sweeps.  Several tests revealed
that increasing the number of relaxation sweeps beyond 50 did not affect
the acceptance rates in a substantial way.  In our implementation, 50
Gau\ss -Seidel sweeps over all slices of a given direction $\tau$
required the same CPU time (on a CRAY Y-MP) as four Creutz heatbath
$SU(2)$ update sweeps.

We checked the validity of the acceptance formula (\ref{formula})
using Monte Carlo estimates for $m_D$ and $P$.
Figure 6 shows results for $\beta = 2.6$ on a $20^4$ lattice.
The results perfectly justify the usage of the approximation formula.
It is therefore sufficient to study the behavior of the quantities $m_D$
and $P$.  Our Monte Carlo results are presented in table \ref{tab3}.
The last column gives the statistics in sweeps (equilibration sweeps are
not counted here).
We used a mixture of four microcanonical
overrelaxtion sweeps followed by a single Creutz heat bath sweep.
Measurements (including the determination of $g$) were performed every
25 sweeps.

\begin{table}
 \centering
 \caption[dummy]{\label{tab3} Monte Carlo results for $m_D$ and $P$ }
 \vspace{2ex}
\begin{tabular}{|c|c|c|c|c|}
\hline\str
 lattice size & $\beta$  &  $m_D$ &  $P$   & statistics
\\[.5ex] \hline \hline \str
 $8^4 $  & $2.4$ & 0.507(2)   &  1.5131(6)   & 10,000
\tabhline
 $12^4$  & $2.4$ & 0.4957(4)  &  1.5121(3)   & 10,000
\tabhline
 $16^4$  & $2.4$ & 0.4955(2)  &  1.5119(1)   & 10,000
\\[.3ex] \hline \hline \str
 $8^4$  & $2.6$ &  0.497(4)  &   1.7429(3)   & 30,000
\tabhline
 $12^4$  & $2.6$ & 0.465(2)   &  1.7424(2)   & 20,000
\tabhline
 $16^4$  & $2.6$ & 0.4644(3)  &  1.7421(1)   & 10,000
\tabhline
 $20^4$  & $2.6$ & 0.4650(2)  &  1.7422(1)   & 5,000
\\[.3ex] \hline
\end{tabular} \end{table}

In table \ref{tab4} we display the ratios of the disorder mass $m_D$
with two physical masses, the square root of the string tension $\kappa$
and the lowest glue ball mass $m_{0^+}$.  The estimates for the physical
masses are taken from ref.\ \cite{michael}.  The results show that the
disorder mass is nearly independent of $\beta$ in the range studied,
whereas the physical masses decrease by roughly a factor of two.
Thus, $m_D$ is not scaling like a physical mass for the couplings
studied here.
We conclude from this that for large blocks the term quadratic in $L_B$
will strongly suppress the acceptance rates even
when the ratio of correlation length and block size $L_B$ is kept
constant.

\begin{table}
 \centering
 \caption[dummy]{\label{tab4} Comparison of $m_D$ with physical masses}
 \vspace{2ex}
\begin{tabular}{|c|c|c|c|c|c|c|}
\hline\str
 lattice size & $\beta$  &  $m_D$ &  $\sqrt{\kappa}$ & $m_{0^+}$ &
 $m_D/\sqrt{\kappa}$ & $m_D/m_{0^+}$
\\[.5ex] \hline \hline \str
 $16^4$  & $2.4$ & $0.4955(2)$  & $0.258(2)$  & $0.94(3)$ &
 $1.92$  & $0.53$
\tabhline
 $20^4$  & $2.6$ & 0.4650(2)  &  0.125(4)   & 0.52(3) &
 $3.72$  & $0.89$
\\[.3ex] \hline
\end{tabular} \end{table}

However, one could hope that the value of the unwanted mass term
is so small that it does no harm in practical calculations.
Let us examine the effect of this mass term in more detail.  Recall that
$h_1$ is built up from two contributions.  The first contribution is
that related to the gauge field disorder inside the block and is
quantitatively represented by the mass $m_D$.  The second contribution
is associated with the block surface.  The latter can of course be
made smaller by using smooth kernels $\psi$ instead of the piecewise
constant kernels discussed so far. However,the disorder
term cannot be expected to become smaller for smooth kernels (see
below).  In figure 7 we plotted separately
the two contributions to $h_1$
\bea
h_{1,A} &=& \frac{3 \beta}{2}
A \, (L_B -1) L_B^2 \, \sin^2(s L_B^{-1/2}/2) \, ,
\nonumber \\
h_{1,P} &=& 3 \beta P
 \, L_B^2 \bigl( 1- \cos(s L_B^{-1/2}/2) \bigr) \, ,
\eea
for $\beta=2.6$ and block size $L_B=8$ on a $20^4$ lattice.  The plot
shows that already for this block size the disorder contribution is by
no means small -- it is comparable to the surface effect.  It
is therefore not clear that one could achieve any significant
improvement by using smooth kernels.  To investigate this in more
detail, we derive an expression for $h_1$, valid for smooth kernels as
well:
\be
h_1 = \frac{3 \beta}8 s^2 A
                     \sum_{x \in \Lamt} \psi_x^2+
  \frac{\beta}{16}s^2 (P-A) \alpha_3 + O(s^4) \, .
\ee
Since $\sum \psi_x^2 \sim L_B^2$, we get essentially the
same behavior for the disorder contribution as in the case
of piecewise constant kernels.

For smooth $\psi^{sine}$ kernels we show separately in figure 8
the two contributions to $h_1$
\be
h_{1,A} = \frac{3 \beta}8 s^2 A
                     \sum_{x \in \Lamt} \psi_x^2 \, ,\;\;\;
h_{1,P-A}= \frac{\beta}{16}s^2 (P-A) \alpha_3  \, ,
\ee
for $\beta=2.6$ and block size $L_B=8$ on a $20^4$ lattice.
We observe that the surface effects are lowered by the smooth kernels,
but the disorder contribution is even higher than for piecewise constant
kernels.

Piecewise constant kernels have the practical feature that once the
change of the Hamiltonian has been calculated, one can perform
multihit Metropolis updating or microcanonical overrelaxation.
In a special case, even an explicit multigrid implementation
with a W-cycle is possible (see below).
For smooth kernels the change in the Hamiltonian would have to
be calculated again and again.  Also the advantage of smooth kernels
are not that clear on small three or four dimensional blocks.  For an
actual simulation we would therefore prefer piecewise constant kernels.

\subsubsection{Maximally abelian gauge}

Our proposal for the choice of $g$ was motivated by the desire to
minimize the quantity $A$.  We now ask whether there is a better choice
of $g$ than the $g$ determined by the Coulomb gauge condition.  For the
sake of simplicity let us take $\vn = (0,0,1)$, i.e.
$\vn \!\cdot\! \vs \, = \sigma_3$.
Then $A$ is given by
\be
 A =  - \bigl\langle \Tr\bigl(( \sigma_3 \, \Um{x}^g \, \sigma_3 -
\Um{x}^g) H_{x,\mu}^{g\, *} \bigr) \bigr\rangle \, .
\ee
The choice of
the Coulomb gauge condition aimed at bringing $\Um{x}^g$ as close to
unity as possible.  Alternatively, one might require that $\Um{x}^g$
should be as close as possible to a $SU(2)$-matrix of the form
$a_0 + ia_3 \sigma_3$. This will also lead to
a small $A$.
The corresponding gauge transformation $g$ can be found by maximizing
the functional
\be\label{abel_gauge_slice} G_A(U,g) =
\sum_{(x,x+\hat\mu) \in \Lamt} \Tr \bigl( \sigma_3 \Um{x}^g \sigma_3
\Um{x}^{g \, *} \bigr) \, ,
\ee
leading to the maximally abelian gauge
\cite{maxabel}, here implemented only on a slice.  We computed $m_D$
also using the $g$'s resulting from this gauge condition and compared
the results with the ones obtained by using the Coulomb gauge condition.
We did not find a substantial difference.  We prefer the Coulomb gauge
condition because it does not depend on the direction $\vn$ and thus
saves computer time.

\subsubsection{Proposal for an implementation}

An explicit multigrid
implementation is possible in a special case
if we use piecewise constant kernels.
This was pointed out in ref.\ \cite{newgauge}
in a related context.

The idea is to update only a fixed $U(1)$ subgroup
of $SU(2)$ globally:
We divide the fundamental lattice $\Lambda_0$ in hypercubic blocks $x'$
of size $2^4$ and ``rotate'' all links going from sites
$x \in x'$ in a fixed $\tau$-direction
with the same angle $\theta_{x'}$:
\be
\Ut{x} \rightarrow \Ut{x}' = g_x^* R_x g_x \, \Ut{x} \, ,
\ee
with
\be
R_x(\vn_{x'},\theta_{x'}) = \cos( \theta_{x'} /2 )
 + i \sin( \theta_{x'} /2) \,
\vn_{x'} \!\cdot\! \vs \, .
\ee
The gauge transformation $g$ is obtained by imposing the
Coulomb gauge condition on slices $\Lamt$ as defined above.
We now consider the special case where the directions $\vn_{x'}$ of the
embedded $U(1)$-subgroups are independent of the block $x'$, i.e.\
$\vn_{x'} = \vn $ for all $x' \in \Lambda_1$.
Then we get a conditional Hamiltonian $\calH_1(\theta)$ by
    substituting
the ``rotated'' gauge field $U'$ in the fundamental Hamiltonian.
By iterating this procedure one gets conditional Hamiltonians
$\calH_k(\theta)$ on coarser layers $\Lambda_k$.
The point is that in the special case considered here
$\calH_k(\theta)$ always stays of the form
\bea
-\calH_k(\theta) &=& \half \sum_{x' \in \Lambda_k}
\left\{
          \beta_{x'}^{cc} \cos^2(\theta_{x'})
+      \beta_{x'}^{cs}\cos(\theta_{x'})\sin(\theta_{x'})\right.
+ \left. \beta_{x'}^{ss}\sin^2(\theta_{x'}) \right\}
\nonumber \\
&+&      \half \sum_{x' \in \Lambda_k}\sum_{\mu \neq \tau}
\left\{
\beta_{x',\mu}^{cc} \cos(\theta_{x'})\cos(\theta_{x'+\hat{\bf \mu}})
\right.
+\beta_{x',\mu}^{cs}\cos(\theta_{x'})\sin(\theta_{x'+\hat{\bf \mu}})
\nonumber \\
&+& \beta_{x',\mu}^{sc}\sin(\theta_{x'})\cos(\theta_{x'+\hat{\bf \mu}})
+ \left.
 \beta_{x',\mu}^{ss}\sin(\theta_{x'})\sin(\theta_{x'+\hat{\bf \mu}})
                                               \right\}
\nonumber \\
&+& const \, ,
\eea
with space dependent couplings that can be recursively calculated from
the couplings defined on the next finer layer $\Lambda_{k-1}$.
Therefore a W-cycle is possible.  Of course, this implementation is
also possible with three dimensional blocks.

\section{Summary}

\label{SECsummary}

We have presented a simple yet accurate formula that
expresses acceptance rates for nonlocal update algorithms in terms of
one single parameter (or two in the case of nonabelian gauge theory)
entering the quantity $h_1$.  This parameter is easy to compute,
e.g.\ by Monte Carlo simulations on a small lattice.  We encountered two
classes of models.  For Sine Gordon, $\phi^4$ theory and $SU(2)$ lattice
gauge theory, $s$ had to be rescaled like $L^{-1}_B$ for piecewise
constant and for smooth kernels, whereas for massless free field theory,
the XY model, the $O(N)$ nonlinear $\sigma$-model and $U(1)$ lattice
gauge theory, one can achieve $L_B$-independent acceptance rates by
choosing smooth kernels.

We can compare the behavior of the acceptance rates in interacting
models with free field theory, where CSD is known to be eliminated by a
multigrid algorithm.  In order to do this we presented a study of the
influence of the coarse-to-fine interpolation on the acceptance rates in
free field theory.

The results of the comparison are consistent with the following rule:
For an interacting
model, sufficiently high acceptance rates for a complete elimination of
CSD can only be expected if $\, h_1 = \langle \calH(\ph+s \psi) -
\calH(\ph) \rangle $ contains no algorithmic ``mass'' term
$\sim s ^2 \sum_x \psi^2_x$.  With the help of this rule it is possible
to decide whether a
given statistical model is a natural candidate for multigrid Monte Carlo
or not.

The kinematical mechanism that leads to a
failure of multigrid algorithms is well described by our analysis.
We hope that a better understanding can lead to improved
multigrid algorithms that can overcome kinematical
obstructions stemming from an algorithmic ``mass'' term.

The acceptance rates of our proposal for nonlocal updates in
$SU(2)$ lattice gauge theory were investigated in detail.  Here we found
that an algorithmic ``mass'' term generated by the disorder in the gauge
field suppresses the acceptance rates on large blocks.  From this study
we do not expect that our algorithm will have a chance to overcome CSD.
However, we believe that compared to local Metropolis algorithms there
will be an acceleration of the Monte Carlo dynamics by a constant factor
(depending on the details of the implementation).  We think that
the best method for practical purposes would be an explicit multigrid
implementation using piecewise constant kernels and a W-cycle. An
implementation and test is planned.
The crucial question is, of course, whether one is able to beat
the overrelaxation algorithm.

\section*{Acknowledgements}

Our study on multigrid Monte Carlo algorithms
for nonabelian gauge theory was inspired by
related work of Thomas Kalkreuter, Gerhard Mack and Steffen Meyer.  We
would like to thank them for many stimulating discussions.  The idea of
our $SU(2)$ algorithm was developed together with Martin Hasenbusch.  It
is a pleasure for us to thank him for the enjoyable collaboration.

Our $SU(2)$ program contains some $SU(2)$ update routines written by
Hans Gerd Evertz and Mihai Marcu.

The numerical computations were performed on the NEC SX-3 in Cologne,
the Landesvektorrechner of the RWTH in Aachen and the CRAY Y-MP of the
HLRZ in J\"ulich.

One of us (M.G.) would like to thank the Deutsche Forschungsgemeinschaft
for financial support.

{\,}

\newpage
{\Large \bf Figure Captions}
\vskip6mm \mbox{}
\\
FIG.1: Metropolis step sizes $\varepsilon(L_B)$
       for massless free field theory in two dimensions,
       $512^2$-lattice.
       $\varepsilon(L_B)$ is choosen in such a way that
       always $P_{acc} = 0.5$ holds.
       Symbols:
       full circles: $\psi^{exact}$,
       full triangles: $\psi^{trunc}$,
       empty circles: $\psi^{min}$,
       empty triangles: $\psi^{sine}$,
       full squares: $\psi^{linear}$,
       empty squares: $\psi^{const}$.
       Lines are only drawn to guide the eye.
\vskip6mm
\noindent
FIG.2: Metropolis step sizes $\varepsilon(L_B)$
       for massless free field theory in four dimensions,
       $64^4$-lattice.
       $\varepsilon(L_B)$ is choosen in such a way that
       always $P_{acc} = 0.5$ holds.
       Symbols:
       full circles: $\psi^{exact}$,
       full triangles: $\psi^{trunc}$,
       empty circles: $\psi^{min}$,
       empty triangles: $\psi^{sine}$,
       full squares: $\psi^{linear}$,
       empty squares: $\psi^{const}$.
       Lines are only drawn to guide the eye.
\vskip6mm
\noindent
FIG.3: $\Omega(s)$ for piecewise constant kernels
       in the 2-dimensional Sine Gordon model, $\beta=39.478$,
       $\zeta=1$.  From top to bottom:
       $L_B=4,8,16,32$ on a $16^2, 32^2, 64^2$, $128^2$ lattice,
       respectively. Points with error bars:  Monte Carlo results,
       lines:  analytical results.
\vskip6mm
\noindent
FIG.4: $\Omega(s)$ for piecewise constant
       kernels in the 2-dimensional XY model,
       $\beta=1.2$.
     {}From top to bottom: $L_B=4,8,16$ on a
       $16^2,32^2,64^2$ lattice, respectively.
       Points with error bars: Monte Carlo results,
       lines: analytical results.
\vskip6mm
\noindent
FIG.5: $\Omega(s)$ for piecewise constant kernels
       in the 2-dimensional $\phi^4$ theory,
       $m_o^2=-0.56$, $\lambda_o=2.4$.
     {}From top to bottom: $L_B=4,8,16$ on a
       $16^2,32^2,64^2$ lattice, respectively.
       Points with error bars: Monte Carlo results,
       lines: analytical results.
\vskip6mm
\noindent
FIG.6: $\Omega(s)$ in 4-dimensional $SU(2)$ lattice
       gauge theory using piecewise constant kernels,
       $\beta=2.6$ on a $20^4$-lattice.
     {}From top to bottom: $L_B=2,4,8,16$.
       Points with error bars: Monte Carlo results,
       lines: analytical results using $m_D$ and $P$
       from Monte Carlo (errors smaller than line width).
\vskip6mm
\noindent
FIG.7: Comparison of disorder and surface effects for
       4-dimensional $SU(2)$ lattice gauge theory using
       piecewise constant kernels on an $8^3$-block,
       $\beta=2.6$ on a $20^4$-lattice.
       Solid line: $h_{1,A}(s)$ (disorder effects),
       dashed line: $h_{1,P}(s)$ (surface effects).
\vskip6mm
\noindent
FIG.8: Comparison of disorder and surface effects for
       the 4-dimensional $SU(2)$ lattice gauge theory using
       smooth $\psi^{sine}$ kernels on a $8^3$-block,
       $\beta=2.6$ on a $20^4$-lattice,
       quadratic approximation used.
       Solid line: $h_{1,A}(s)$ (disorder effects),
       dashed line: $h_{1,P-A}(s)$ (surface effects).
\end{document}